\documentclass{PoS}
\usepackage{graphicx}
\usepackage{epstopdf}
\usepackage{amsmath}
\usepackage{multirow}
\usepackage{array}
\title{Antibaryon interactions with the nuclear medium}

\ShortTitle{Antibaryon interactions with the nuclear medium}

\author{\speaker{Jaroslava Hrt\'{a}nkov\'{a}}, Ji\v{r}\'{i} Mare\v{s}\\
        Nuclear Physics Institute, 250 68 \v{R}e\v{z}, Czech Republic\\
        E-mail: \email{hrtankova@ujf.cas.cz}}

\abstract{This contribution deals with our recent study of antibaryon interactions with the nuclear medium within the relativistic mean-field approach using antibaryon coupling constants consistent with available experimental data. We performed calculations of $\bar{B}$ ($\bar{B}=\bar{p}, \bar{\Lambda}, \bar{\Sigma}, \bar{\Xi}$) bound states in selected nuclei. Due to the lack of information on the in-medium antihyperon annihilation near threshold only the $\bar{p}$ absorption was considered. It was described by the imaginary part of a phenomenological optical potential fitted to $\bar{p}$-atom data. The annihilation was treated dynamically, taking into account explicitly the reduced phase space for annihilation products in the nuclear medium, as well as the compressed nuclear density due to the antiproton. The energy available for the annihilation products was evaluated self-consistently, considering additional energy shift due to particle momenta in the ${\bar p}$-nucleus system. Corresponding $\bar{p}$ widths were significantly reduced, however, they still remain sizable.
Next, the $\bar{p}$-nucleus interaction was constructed using the latest version of the Paris $\bar{N}N$ potential. Related scattering amplitudes were used to define the complex $\bar{p}$ optical potential in the nuclear medium. The resulting $\bar{p}$ $1s$ binding energies are about 10\% smaller and widths about 20\% larger than those obtained with the phenomenological approach.}

\FullConference{The 26th International Nuclear Physics Conference\\
		11-16 September, 2016\\
		Adelaide, Australia}

\begin{document}

\section{Introduction}

The antibaryon--nucleus interaction is an interesting and topical issue in view of the future experiments at FAIR facility. Its study could provide us with information about the behavior of an antibaryon inside the medium as well as nuclear dynamics. Moreover, it could serve as a test for models of (anti)hadron--hadron interactions. In particular, much attention was devoted to the $\bar{p}$-nucleus interaction and possible existence of $\bar{p}$-nuclear quasi-bound states \cite{Mishustin}. It was argued in Ref.~\cite{Mishustin} that the phase space for $\bar{p}$ annihilation products in the medium could be substantially suppressed so that $\bar{p}$ could live relatively long inside the nucleus.

In this contribution, we report on our recent self-consistent calculations of $\bar{B}$ bound states in various nuclei using G-parity motivated coupling constants. Special attention was devoted to calculations of $\bar{p}$-nuclear bound states, which were performed using a phenomenological optical potential as well as microscopic Paris $\bar{N}N$ potential. 

In Section 2, we briefly introduce the model used in our calculations. Our results are presented in Section 3 and conclusions are drawn in Section 4.

\section{Model}
The interactions of an antibaryon with $A$ nucleons are studied within the relativistic mean-field approach (RMF) \cite{Walecka}. In this model, the (anti)baryons interact among each other by the exchange of the scalar ($\sigma$)
 and vector ($\omega_{\mu}$, $\vec{\rho}_\mu$) meson fields, and the massless photon field $A_{\mu}$. The equations of motion are derived from the standard Lagrangian density $\mathcal{L}_N$ extended by the Lagrangian density $\mathcal{L}_{\bar{B}}$ describing the antibaryon 
interaction with the nuclear medium using the variational principle (see Ref.~\cite{jarka} for details). The Dirac equations for nucleons and antibaryon read:  
\begin{equation} \label{Dirac antiproton}
[-i\vec{\alpha}\vec{\nabla} +\beta(m_j + S_j) + V_j]\psi_j^{\alpha}=\epsilon_j^{\alpha} \psi_j^{\alpha}, 
\quad j=N,\bar{B}~,
\end{equation}
where
\begin{equation}
S_j=g_{\sigma j}\sigma, \quad V_j=g_{\omega j} \omega_0 + g_{\rho j}\rho_0 \tau_3 + e_j \frac{1+\tau_3}{2}A_0
\end{equation}
are the scalar and vector potentials, respectively. Here, $\alpha$ denotes single particle states, $m_j$ stands for 
(anti)baryon masses and $g_{\sigma j}, g_{\omega j}, g_{\rho j}$, and $e_j$ are (anti)baryon coupling 
constants to corresponding fields. The Klein--Gordon equations for the meson fields involve additional source terms due to the antibaryon:
\begin{equation}
\begin{split} \label{meson eq}
(-\triangle + m_{\sigma^2}+ g_2 \sigma + g_3 \sigma^2) \sigma&=- g_{\sigma N} \rho_{SN}-g_{\sigma \bar{B}} 
\rho_{S \bar{B}}~, \\
(-\triangle + m_{\omega^2} +d \omega^2_0) \omega_0&= g_{\omega N}\rho_{VN} +g_{\omega \bar{B}} \rho_{V\bar{B}}~, \\
(-\triangle + m_{\rho^2}) \rho_0&= g_{\rho N}\rho_{IN} +g_{\rho \bar{B}}\rho_{I \bar{B}}~, \\
-\triangle A_0&= e_N \rho_{QN}+e_{\bar{B}}\rho_{Q\bar{B}}~,
\end{split}
\end{equation}
where $m_\sigma, m_\omega, m_\rho$ are the masses of
considered mesons and $\rho_{\text{S}j}, \rho_{\text{V}j}, \rho_{\text{I}j}$ and $\rho_{\text{Q}j}$  are the scalar, 
vector, isovector and charge densities, respectively. The system of coupled Dirac  \eqref{Dirac antiproton}  and Klein--Gordon  
\eqref{meson eq} equations is solved self-consistently by iterative procedure.

The values of the nucleon--meson coupling constants and meson masses were adopted from the nonlinear RMF 
models TM1(2) \cite{Toki} for heavy (light) nuclei and from the NL-SH model \cite{nlsh}. The hyperon--meson 
coupling constants for the $\omega$ and $\rho$ fields were derived using SU(6) symmetry relations. The values of the $\sigma$ coupling constants were obtained from fits to available experimental data --- $\Lambda$ hypernuclei 
\cite{hyperon couplings}, $\Sigma$ atoms \cite{sigma atomy}, and $\Xi$ production in $(K^+, K^-)$ reaction 
\cite{Khaustov}.

The $\bar{B}$--nucleus interaction is constructed from the $B$--nucleus interaction with the help of the 
G-parity transformation: the potential generated by the exchange of the $\omega$ meson changes sign due to the G-parity and becomes attractive. The G-parity is surely a valid concept for the long and medium range $\bar{B}$ potential. It yields a very deep $\bar{B}$-nucleus potential, e. g., the $\bar{p}$ potential would be about $750$~MeV deep inside a nucleus. However, the $\bar{B}$ annihilation, which is a dominant process in the short range interaction, and various many-body effects could cause significant deviations from the G-parity values
in the nuclear medium. Indeed, the experiments with antiprotonic atoms~\cite{mares} and $\bar{p}$ scattering off nuclei at low energies \cite{antiNN interaction} suggest that the real part of the $\bar{p}$-nucleus potential is $100 - 300$~MeV deep in the nuclear interior. Therefore, we introduce a scaling factor $\xi$ for the antibaryon--meson 
coupling constants which are in the following relation to the baryon--meson couplings:
\begin{equation} \label{reduced couplings}
g_{\sigma \bar{B}}=\xi\, g_{\sigma N}, \quad g_{\omega \bar{B}}=-\xi\, g_{\omega N}, \quad g_{\rho 
\bar{B}}=\xi\, g_{\rho N}~.
\end{equation}
In this work, we consider the value of $\xi=0.2-0.3$ which is in accordance with the experimental data fits. We assume the same scaling for antihyperons, as well, due to the lack of experimental information on antihyperon interactions.

\begin{figure}[t]
\begin{center}
\includegraphics[width=0.6\textwidth]{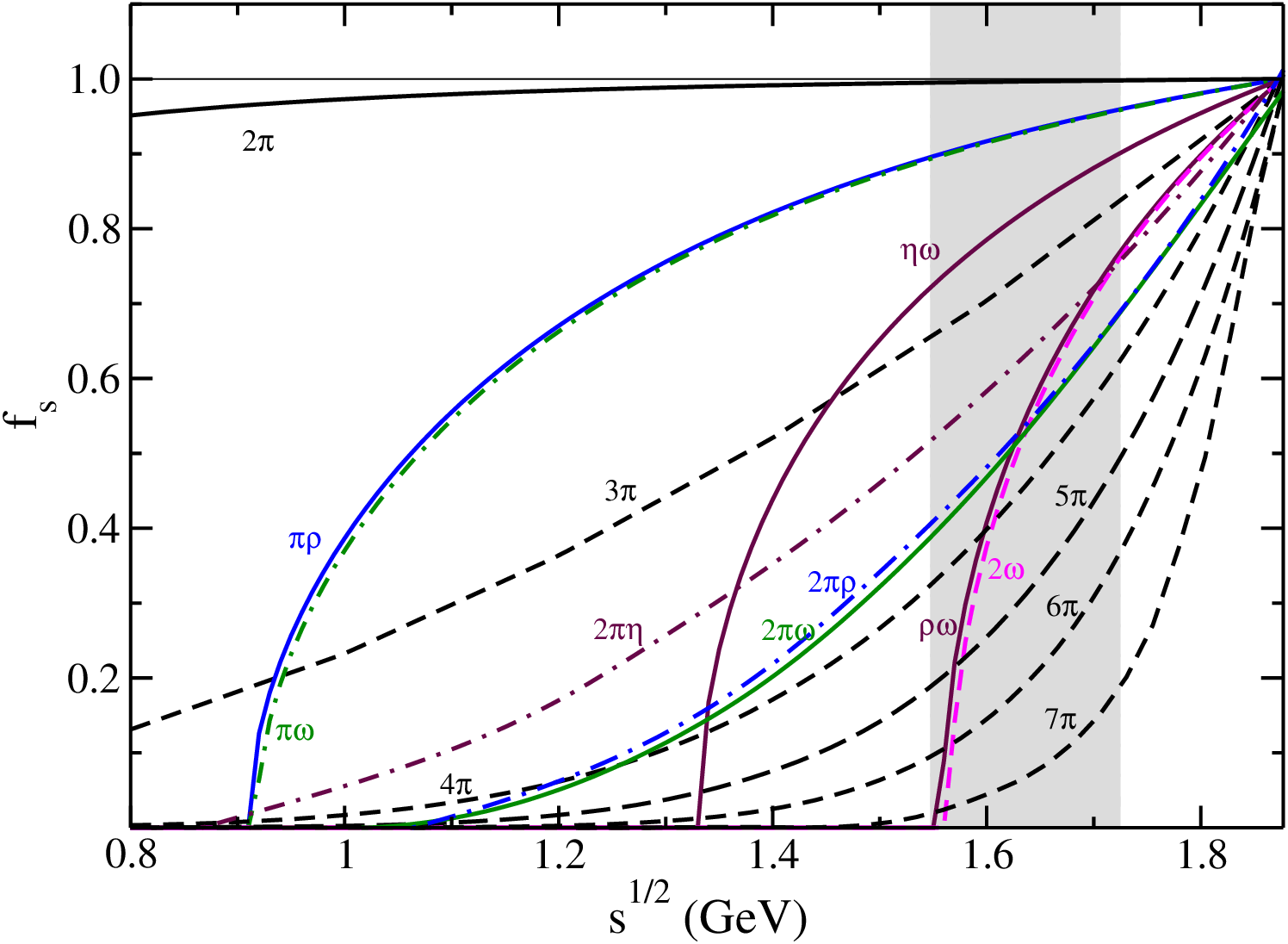}
\end{center}
\caption{The phase space suppression factor $f_s$ as a function of the center-of-mass energy $\sqrt{s}$.}
\label{Fig.:suppF}
\end{figure}

The realistic description of $\bar{B}$-nucleus interaction should involve $\bar{B}$ absorption in the medium. In our calculations, only the $\bar{p}$ absorption 
in a nucleus has been considered since we found no experimental information on antihyperon annihilation in the medium.
The $\bar{p}$ absorption is described by the imaginary part of the optical potential in a `$t\rho$' 
form adopted from optical model phenomenology \cite{mares}:
\begin{equation}
2\mu {\rm Im}V_{\text{opt}}(r)=-4 \pi \left(1+ \frac{\mu}{m_N}\frac{A-1}{A} 
\right){\rm Im}b_0 \rho(r)~,
\end{equation}
where $\mu$ is the $\bar{p}$--nucleus reduced mass. The density $\rho(r)$ is evaluated dynamically 
within the RMF model, while the parameter Im$b_0=1.9$~fm is determined by 
fitting the $\bar{p}$ atom data \cite{mares}. The effective scattering length Im$b_0$ describes the $\bar{p}$ absorption at threshold and, therefore, we evaluate the suppression factor $f_s$ for a given decay channel to account for reduction of the phase space available for decay products of the $\bar{p}$ annihilation in the nuclear medium. The absorptive $\bar{p}$ potential then acquires the form
\begin{equation} \label{im_pot}
 {\rm Im}V_{\bar{p}} (r,\sqrt{s},\rho)=\sum_{\text{channel}} B_c f_{\text{s}}(\sqrt{s}) {\rm Im}V_{\text{opt}}(r)~,
\end{equation}
where $B_c$ is the branching ratio for a given channel (see Ref. \cite{jarka} for details). The calculated phase space suppression factors as a function of $\sqrt{s}$ for all channels considered are depicted in Figure~\ref{Fig.:suppF}.

\begin{figure}[t]
\begin{center}
\includegraphics[width=0.35\textwidth]{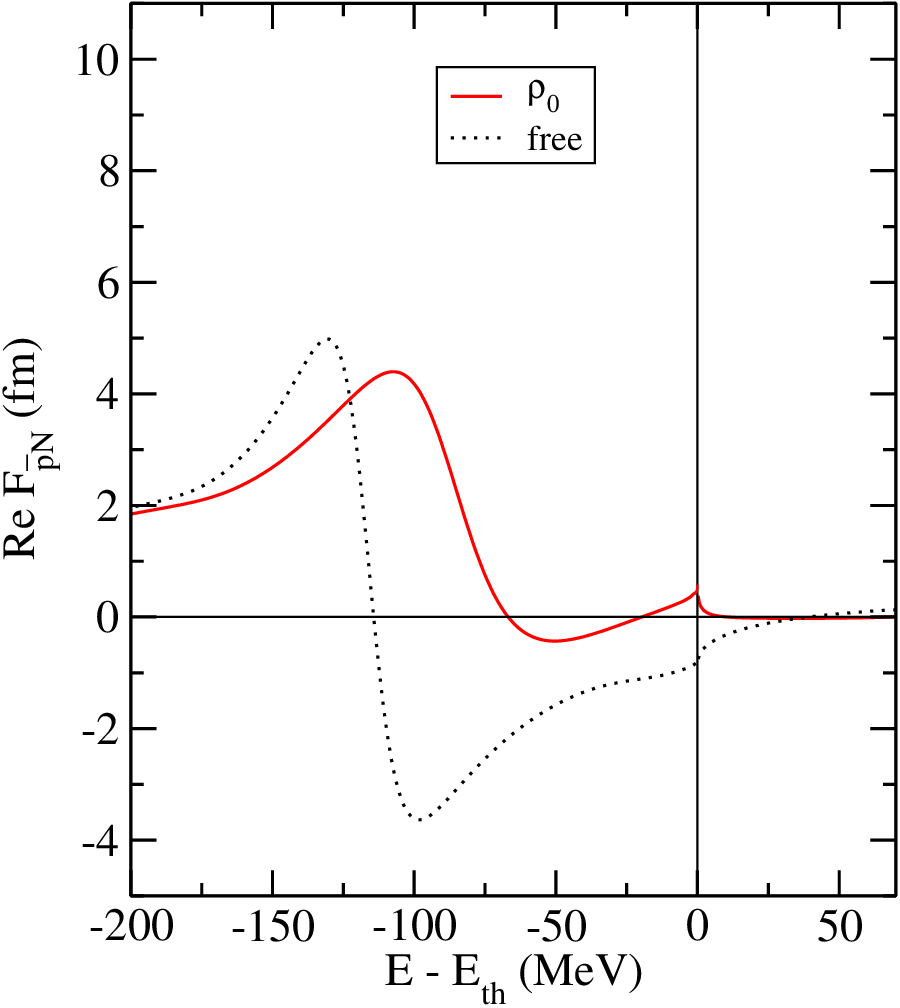} \hspace{10pt}
\includegraphics[width=0.35\textwidth]{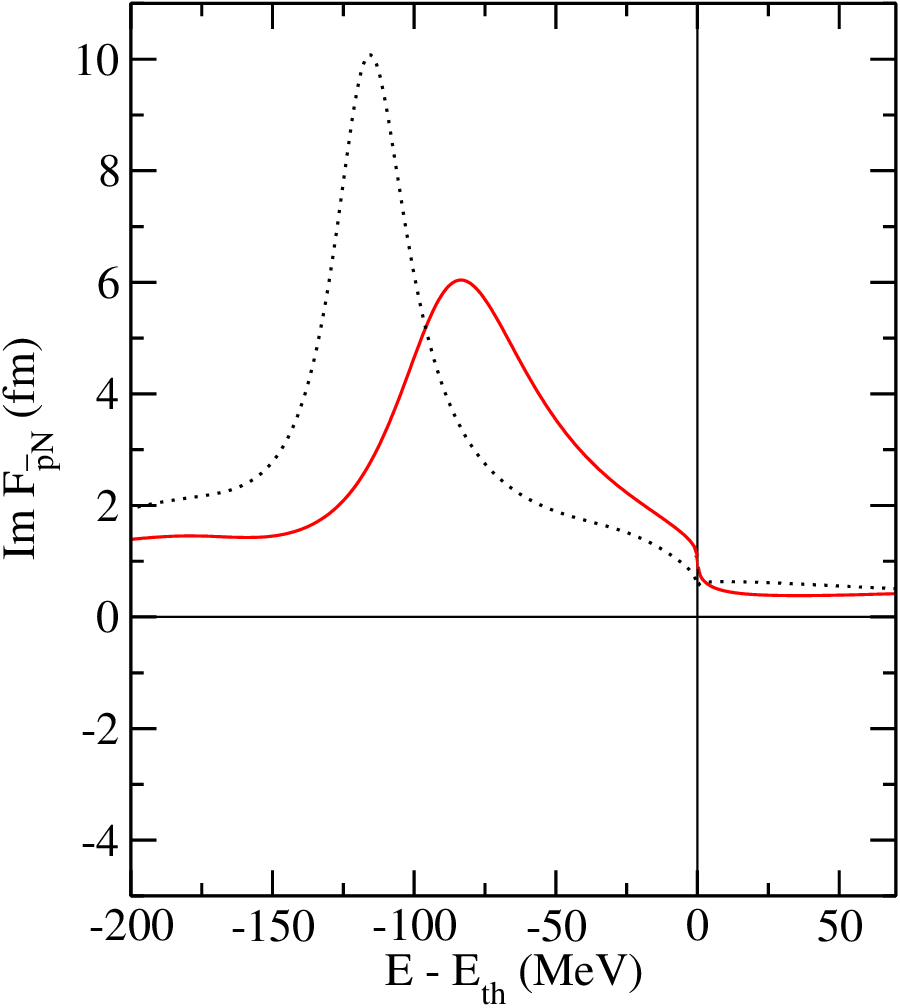}
\end{center}
\caption{Energy dependence of the Paris 09 $\bar{p}N$ S-wave amplitudes: Pauli blocked amplitude for $\rho_0=0.17$~fm$^{-3}$ (solid lines) is compared with free-space amplitude (dotted lines).}
\label{Fig.:amp}
\end{figure}
Next, we construct the $\bar{p}$ optical potential using the S-wave $\bar{p}N$ scattering amplitudes derived from the latest version of the 
Paris $\bar{N}N$ potential~\cite{Paris2}. 
The free-space amplitudes are modified using the multiple scattering approach of Wass et al. \cite{wrw} to account for Pauli correlations in the medium. The in-medium isospin 1 and 0 amplitudes are of the form
\begin{equation}
F_{1}=\frac{f_{\bar{p}n}(\delta \sqrt{s})}{1+\frac{1}{4}\xi_k \frac{\sqrt{s}}{m_N} f_{\bar{p}n}(\delta \sqrt{s}) \rho}~, \qquad F_{0}=\frac{[2f_{\bar{p}p}(\delta \sqrt{s})-f_{\bar{p}n}(\delta \sqrt{s})]}{1+\frac{1}{4}\xi_k \frac{\sqrt{s}}{m_N}[2f_{\bar{p}p}(\delta \sqrt{s}) - f_{\bar{p}n}(\delta \sqrt{s})] \rho}~. 
\end{equation} 
Here, $f_{\bar{p}n}$ and $f_{\bar{p}p}$ denote the free-space amplitudes as a function of $\delta \sqrt{s}=\sqrt{s}-E_{\text{th}}$; 
$\rho$ is the nuclear core density distribution and $\xi_k$ is taken from Ref.~\cite{fgNPA2017}. In Figure~\ref{Fig.:amp}, there are free-space $\bar{p}N$ amplitudes compared with the in-medium modified amplitudes at $\rho_0$ as a function of energy. Both amplitudes vary significantly with energy below threshold. The peaks of the in-medium amplitudes are lower in comparison with the free-space amplitudes and are shifted towards threshold.  
The S-wave optical potential is of the following form:
\begin{equation} \label{parisS}
	2E_{\bar{p}}V_{\text{opt}}=-4\pi \frac{\sqrt{s}}{m_N}\left(F_0\frac{1}{2}\rho_p + F_1\left(\frac{1}{2}\rho_p+\rho_n\right)\right)~,
\end{equation}
  where $\rho_p$ ($\rho_n$) is the proton (neutron) density distribution and the factor 
$\sqrt{s}/m_N$ transforms the in-medium amplitudes to the $\bar{p}$-nucleus frame.
\bigskip

The energy relevant for the $\bar{p}$ scattering amplitudes and suppression factors in the nuclear medium is defined by Mandelstam variable
\begin{equation} \label{s}
s=(E_N + E_{\bar{p}})^2 - (\vec{p}_N + \vec{p}_{\bar{p}})^2~,
\end{equation} 
where $E_N=m_N - B_{Nav}$, $E_{\bar{p}}=m_{\bar{p}}-B_{\bar{p}}$, $B_{Nav}$ and $B_{\bar{p}}$ are the average 
binding energy per nucleon and the $\bar{p}$ binding energy, respectively.  
In the two-body c.m. frame $\vec{p}_N + \vec{p}_{\bar{p}} = 0$ and Eq.~\eqref{s} reduces to
\begin{equation} \label{Eq.:M}
\sqrt{s}=~m_{\bar{p}}+m_{N}-B_{\bar{p}}-B_{Nav}~~~(\text{M}).
\end{equation}
However, in the $\bar{p}$-nucleus frame the momentum dependent term in Eq.~\eqref{s}
is no longer negligible~\cite{s} and provides additional downward energy shift. Then the Mandelstam variable can be rewritten as \begin{equation} \label{Eq.:J}
 \sqrt{s}= E_{th} \left(\!1-\frac{2(B_{\bar{p}} + B_{Nav})}{E_{th}} + \frac{(B_{\bar{p}}+ B_{Nav})^2}{E_{th}^2} - \frac{1}{E_{th}}T_{\bar{p}} - \frac{1}{E_{th}}T_{Nav} \!\right)^{1/2}~~~(\text{J}),
\end{equation}
where $T_{Nav}$ is the average kinetic energy per nucleon 
and $T_{\bar{p}}$ represents the $\bar{p}$ kinetic energy. The kinetic energies were calculated as
the expectation values of the kinetic energy operator ${T}_j=-\frac{\hbar^2}{2 m_j^{(*)}} \triangle$, where $m^*_j=m_j-S_j$ is the
(anti)nucleon reduced mass.

\begin{figure}[b]
\begin{center}
\includegraphics[width=0.52\textwidth]{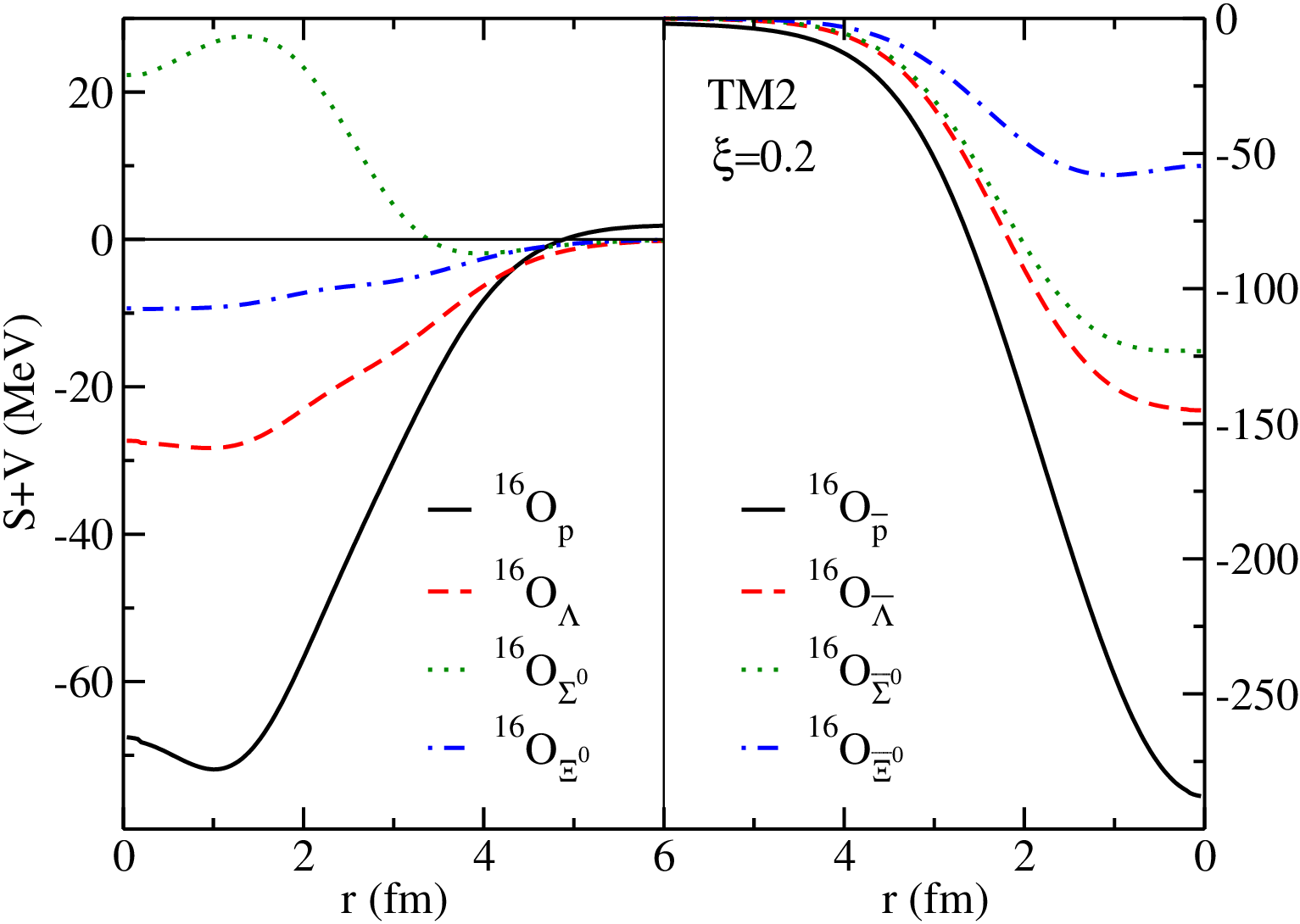}
\end{center}
\caption{The $B$--nucleus (left) and $\bar{B}$--nucleus (right) potentials in $^{16}$O, calculated dynamically in the TM2 model for $\xi=0.2$.}
\label{Fig.:Bpot}
\end{figure}

\section{Results}

First, we performed self-consistent calculations of $1s$ $\bar{B}$ bound states in various nuclei using the RMF model with G-parity motivated coupling constants, introduced in previous section. Then, we considered the $\bar{p}$ absorption inside the nucleus. The $\bar{p}$ absorption was described by the imaginary part of the phenomenological optical potential. Finally, we studied $\bar{p}$ quasi-bound states within the latest version of the Paris $\bar{N}N$ potential. 

\begin{figure}[t]
\begin{center}
\includegraphics[width=0.52\textwidth]{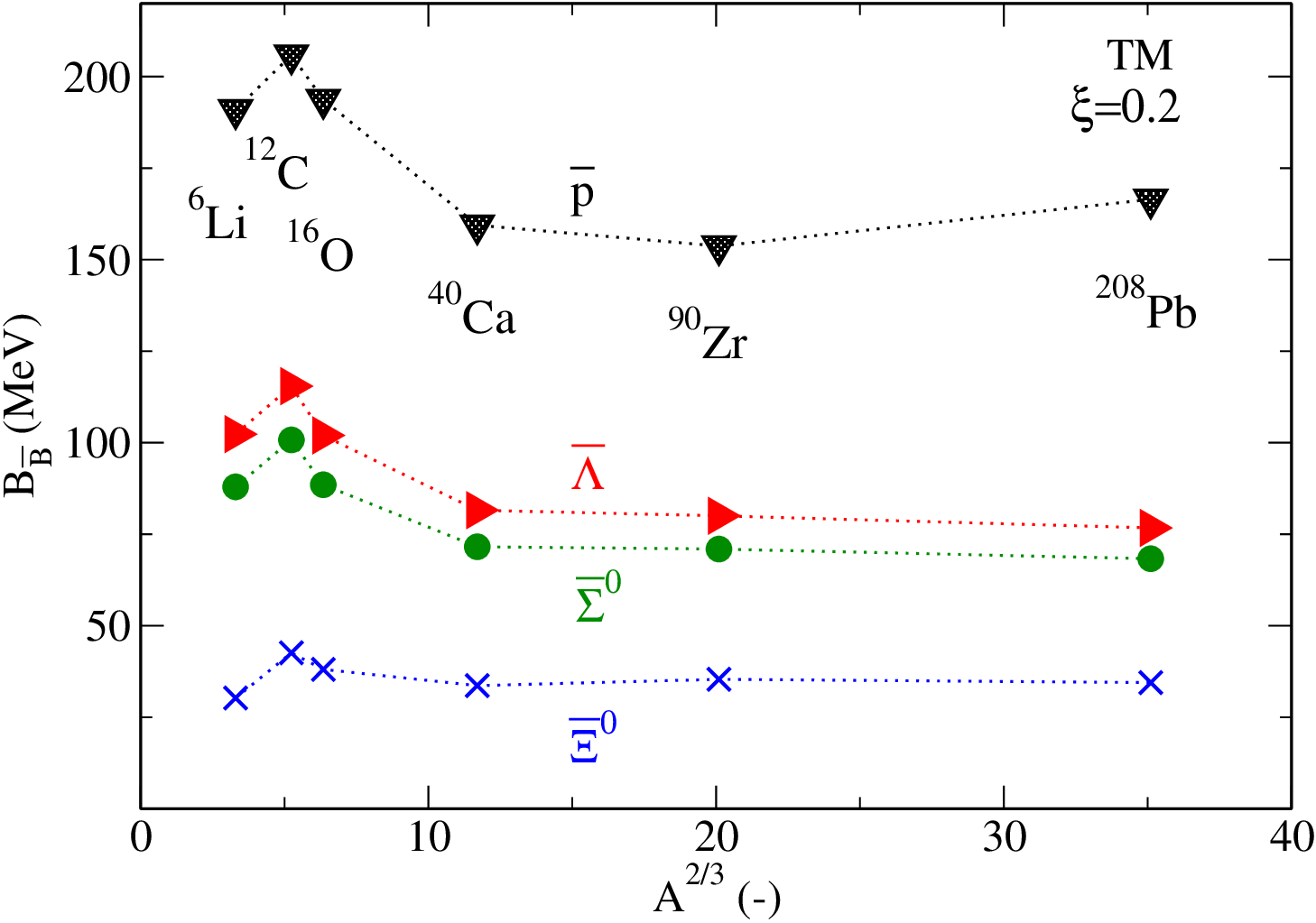}
\end{center}
\caption{The $A$ dependence of $\bar{B}$ $1s$ binding energies, calculated dynamically in the TM model for $\xi=0.2$.}
\label{Fig.:Bene}
\end{figure}

\begin{figure}[b]
\begin{center}
\includegraphics[width=0.35\textwidth]{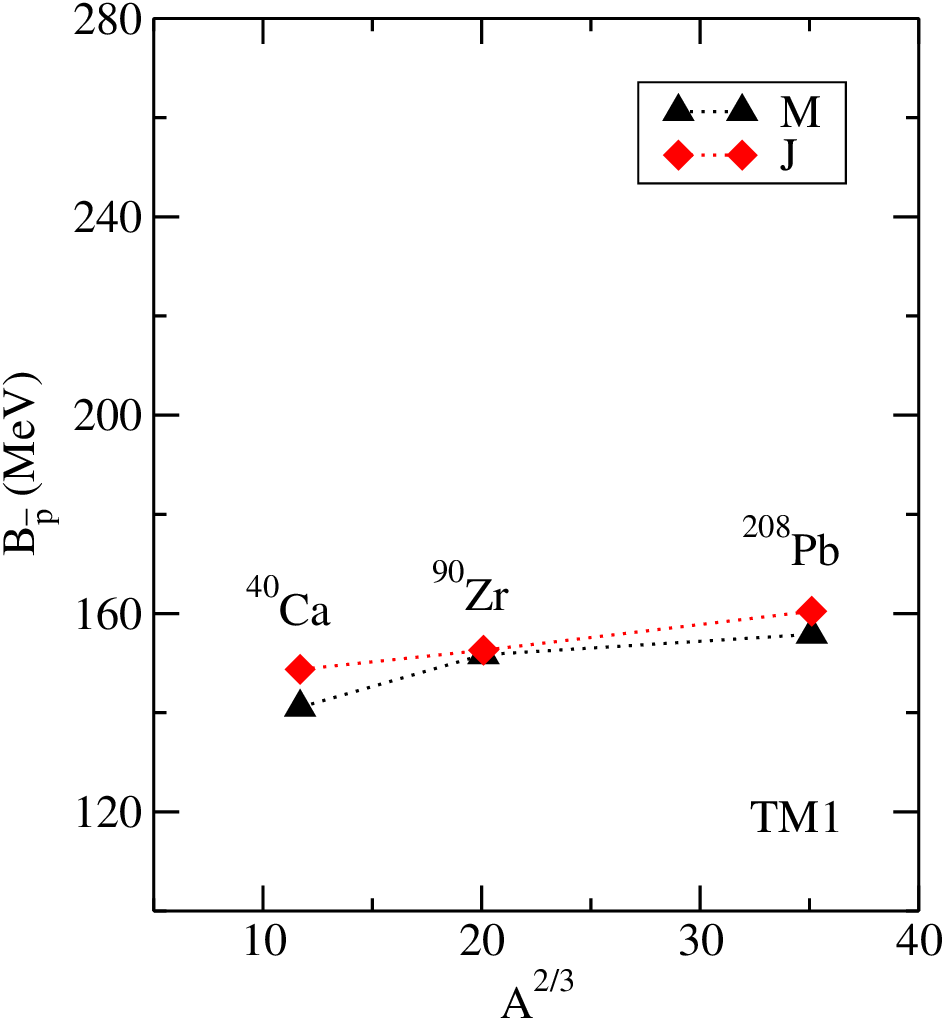} \hspace{10pt}
\includegraphics[width=0.35\textwidth]{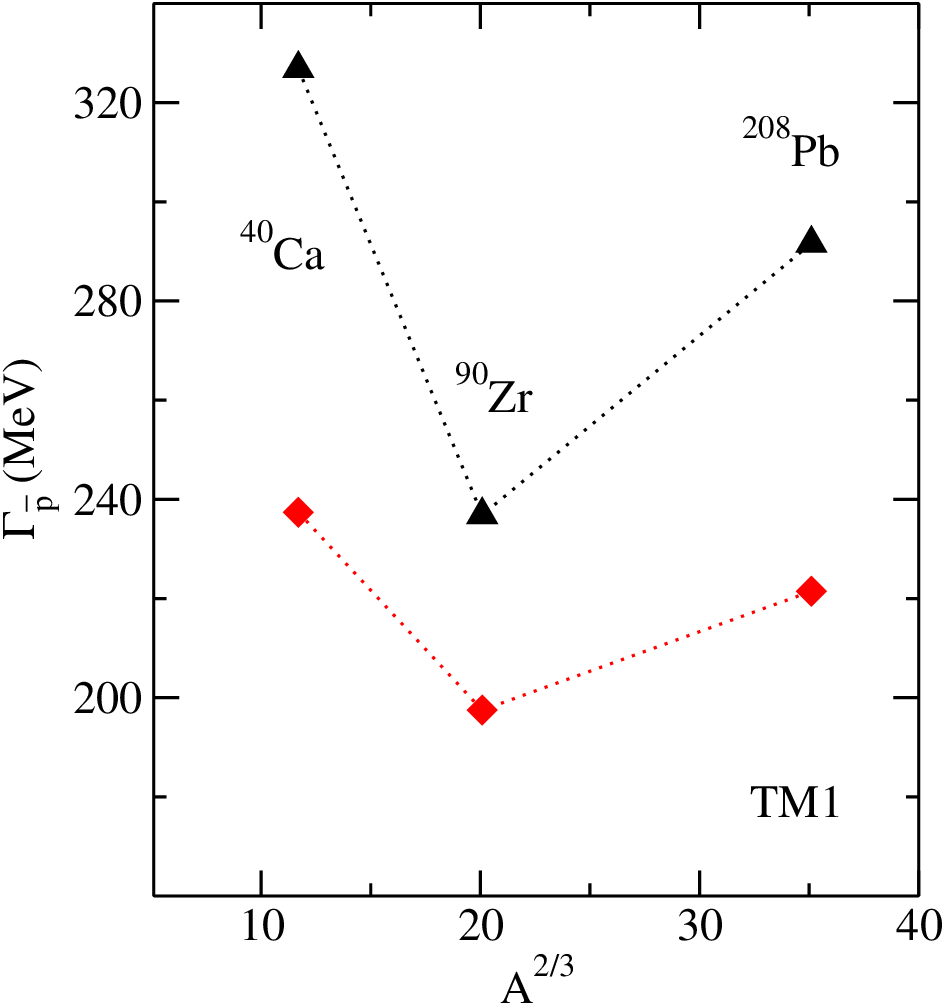}
\end{center}
\caption{Binding energies (left panel) and widths (right panel) of $1s$ $\bar{p}$-nuclear states in selected nuclei,
calculated dynamically using the TM1 model for different $\sqrt{s}$.}
\label{Fig.:Pene}
\end{figure}

In Figure~\ref{Fig.:Bpot}, there is the total potential acting on an extra baryon and extra antibaryon in the $1s$ state in $^{16}$O, calculated dynamically (i. e., the core polarization effect due to $\bar{B}$ was considered) in the TM2 model. All antibaryons feel attractive potential due to the G-parity transformation (note that even $\bar{\Sigma}_0$ feels attraction inside the nucleus). The depth of the potential felt by $\bar{B}$ is deeper than the one felt by $B$ inside the nucleus and indicates that the antibaryons would be strongly bound in the medium. Figure~\ref{Fig.:Bene} presents corresponding $1s$ binding energies of $\bar{B}$ bound in nuclei across the periodic table, calculated dynamically in the TM model and $\xi=0.2$. The $\bar{p}$ is the most bound antibaryon in all nuclei considered since it feels the deepest potential inside the medium. The $\bar{\Lambda},~\bar{\Sigma}_0$ and $\bar{\Xi}_0$ are bound less due to the weaker couplings to the meson fields. It is to be noted that the presented binding energies were calculated in two models, the TM2 model for $^{6}$Li, $^{12}$C and $^{16}$O and the TM1 model for $^{40}$Ca, $^{90}$Zr and $^{208}$Pb. These two models yield different values of nuclear compressibility and different magnitudes of the $\sigma$ and $\omega$ fields and, therefore, the binding energies do not grow with the increasing mass number $A$ as would be expected (see Ref.~\cite{jarka} for more details). 

Next, we considered the $\bar{p}$ absorption inside the nucleus by adding the imaginary part of the phenomenological potential to the real $\bar{p}$-nucleus potential evaluated within the RMF approach. In Figure~\ref{Fig.:Pene}, there are binding energies (left panel) and widths (right panel) of the $1s$ $\bar{p}$-nuclear states in various nuclei, calculated dynamically in the TM1 model. The presented results were calculated for $\sqrt{s}$ in the two-body frame (M) and laboratory frame (J). The two versions of $\sqrt{s}$ yield similar $\bar{p}$ binding energies.
The energies in a given nucleus are not much affected by the $\bar{p}$ absorption (compare with Figure~\ref{Fig.:Bene}). On the other hand, the ${\bar p}$ widths are sizable in the two-body c.m. frame and are significantly reduced after including the momentum dependent term in $\sqrt{s}$. However, they still remain large.

\begin{figure}[t]
\begin{center}
\includegraphics[width=0.35\textwidth]{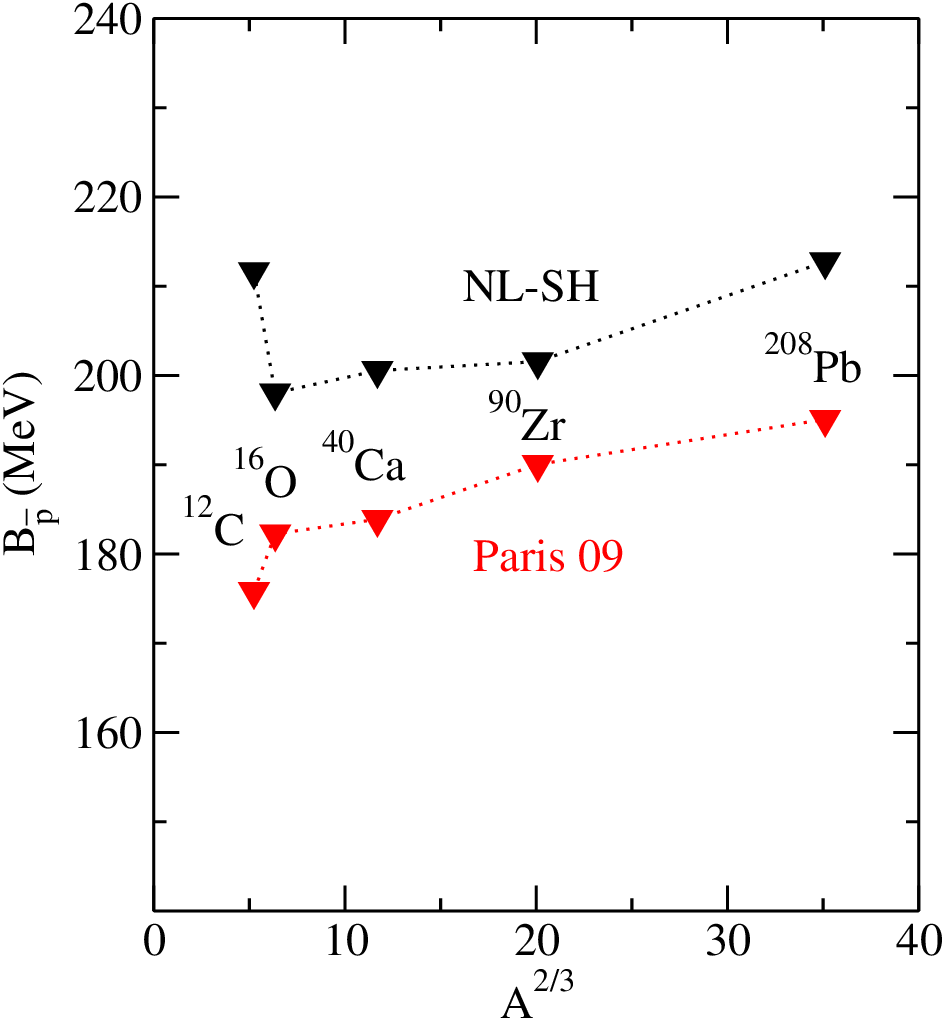} \hspace{10pt}
\includegraphics[width=0.35\textwidth]{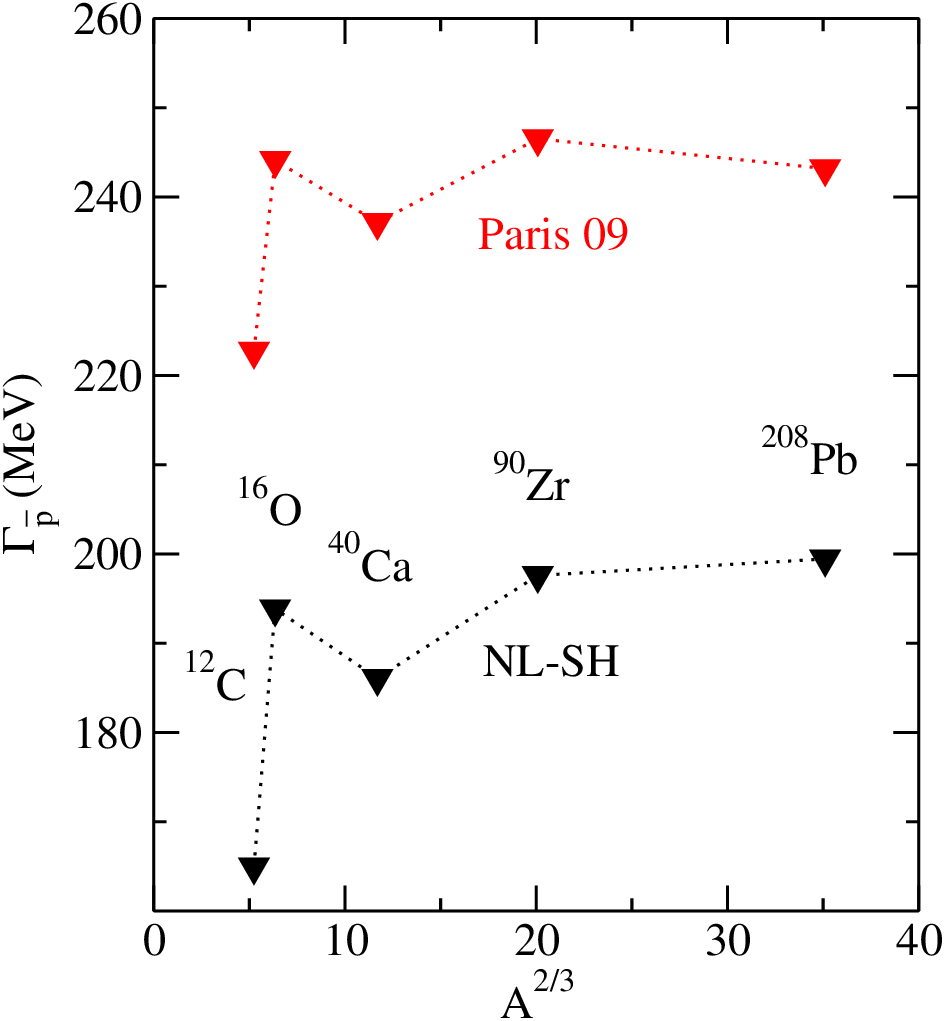}
\end{center}
\caption{Binding energies (left panel) and widths (right panel) of $1s$ $\bar{p}$-nuclear states in selected nuclei,
calculated dynamically for $\sqrt{s}=$ J using the Paris $\bar{N}N$ S-wave potential (red) and phenomenological approach within the NL-SH model (black).}
\label{Fig.:paris}
\end{figure}

We performed a comparable study of $\bar{p}$-nuclear quasi-bound states using the microscopic Paris $\bar{N}N$ S-wave potential. The resulting $1s$ $\bar{p}$ binding energies and corresponding widths are presented in Figure~\ref{Fig.:paris}. The $\bar{p}$ binding energies and widths calculated using the phenomenological approach within the NL-SH model are shown for comparison. The Paris S-wave potential yield smaller $\bar{p}$ binding energies than the phenomenological potential in all nuclei considered. The $\bar{p}$ widths exhibit the same $A$ dependence, however, they are much larger than those calculated with the phenomenological potential. It is to be noted that the Paris $\bar{N}N$ potential contains sizable P-wave interaction which should be included in the calculations. Such calculations have been performed recently and will be published elsewhere.

\section{Conclusions}

We performed self-consistent calculations of antibaryon-nucleus bound states in selected nuclei. First, the $\bar{B}$-nucleus potential was constructed within the RMF approach using the G-parity motivated coupling constants properly scaled to fit available experimental data. The real parts of the potentials felt by $\bar{B}$ inside nuclei are attractive and fairly deep due to the G-parity transformation. In our calculations, we considered only the $\bar{p}$ absorption inside the nucleus so far. The absorption was described by the imaginary part of the phenomenological potential. The phase space suppression factors entering the phenomenological potential were evaluated self-consistently using $\sqrt{s}$ for the two-body frame and $\bar{p}$-nucleus frame. It was found that the energy shift due to $N$ and $\bar{p}$ momenta significantly reduces the $\bar{p}$ widths. However, they still remain sizable for potentials consistent with $\bar{p}$-atom data. Next, we performed calculations of $\bar{p}$-nuclear quasi-bound states using the optical potential constructed from the Paris $\bar{N}N$ S-wave scattering amplitudes. The free-space $\bar{p}N$ amplitudes were modified in order to account for Pauli correlations in medium. The resulting $1s$ $\bar{p}$ binding energies are about 10\% smaller and widths about 20\% larger than those calculated with the phenomenological approach.

\section*{Acknowledgments}
We wish to thank E. Friedman, A. Gal and S. Wycech for valuable discussions, and B. Loiseau for providing us with the $\bar{N}N$ amplitudes. This work was supported by the GACR Grant No. P203/15/04301S.

\end{document}